\documentclass[seceq]{ptptex}

\usepackage{graphicx}
\usepackage{amsmath,amsfonts,bm,amssymb}

\newcommand{\be}{\begin{equation}}
\newcommand{\ee}{\end{equation}}
\newcommand{\bea}{\begin{eqnarray}}
\newcommand{\eea}{\end{eqnarray}}

\newcommand{\eq}[1]{Eq.~(\ref{eq:#1})}
\newcommand{\sect}[1]{Sec.~\ref{sec:#1}}

\newcommand{\calO}{\cal O}
\newcommand{\eT}{\epsilon_T}
\newcommand{\del}{\partial}
\newcommand{\bareta}{\bar{\eta}}

\title{
Critical phenomena in the AdS/CFT duality%
}

\author{
Makoto \textsc{Natsuume}\footnote{E-mail: makoto.natsuume@kek.jp}%
}

\inst{
KEK Theory Center, Institute of Particle and Nuclear Studies, \\
High Energy Accelerator Research Organization,\\
Tsukuba, Ibaraki, 305-0801, Japan
}

\abst{
We review black holes with second-order phase transition in string theory (R-charged black holes and holographic superconductors) and review their static and dynamic critical phenomena. Holographic superconductors have conventional mean-field values for static critical exponents, but R-charged black holes have unconventional ones. For dynamic universality class, holographic superconductors belong to model A, and R-charged black holes belong to model B in the classification of Hohenberg and Halperin. The analysis suggests that some black holes do obey the theory of critical phenomena in condensed matter physics. 
}

\begin{document}

\maketitle


\section{Introduction} \label{sec:intro}

The AdS/CFT duality \cite{Maldacena:1997re,Witten:1998qj,Witten:1998zw,Gubser:1998bc} claims that a gauge theory is equivalent to a gravity system. According to AdS/CFT, non-Abelian plasmas at strong coupling have a very small $\bareta/s$, where $\bareta$ is the shear viscosity%
\footnote{We use $\bareta$ to avoid confusion with the static critical exponent $\eta$. See \sect{critical}.}
and $s$ is the entropy density:
\be
\frac{\bareta}{s} = \frac{1}{4\pi}~.
\ee
It turns out that the prediction is very close to the value of the real quark-gluon plasma at RHIC experiments.%
\footnote{See, {\it e.g.}, Refs.~\citen{Natsuume:2007qq,Natsuume:2008ha} for reviews.}
Partly inspired by this ``success," many people try to apply AdS/CFT to the other real-world systems such as condensed-matter physics. I will describe one such attempt of ours \cite{Maeda:2009wv,Maeda:2008hn}.%
\footnote{Recently, there appeared various papers which study similar issues \cite{Jain:2009uj,Franco:2009if,Aprile:2009ai,Buchel:2009mf,Iqbal:2010eh,Herzog:2010vz,Buchel:2010gd}.} 
Our aim is to study critical phenomena in the context of AdS/CFT. 

In the second-order phase transition, the correlation length $\xi$ diverges and as a result physical quantities also diverge. But there are two kinds of critical phenomena: 
\smallskip
\begin{enumerate}

\item First one is the standard {\it static critical phenomena}. In the static critical phenomena, thermodynamic quantities diverge as powers of the correlation length. The power is a {\it critical exponent}: it is determined from generic properties of the system such as the symmetry, spatial dimensionality, and so on, but not the other details of microscopic physics. This is the universality. 

\item Second one is the {\it dynamic critical phenomena}. In the dynamic case, one interesting quantity is the relaxation time $\tau$. The relaxation time measures how some disturbance decays in time. In the dynamic critical phenomena, the relaxation time also diverges which is known as the {\it critical slowing down}. 

\end{enumerate}
\smallskip
We study these phenomena from gravity side. Our goal is to see if AdS black holes obey the theory of critical phenomena. 
I will argue that holographic superconductors \cite{Gubser:2008px,Hartnoll:2008vx,Hartnoll:2008kx} and R-charged black holes \cite{Behrndt:1998jd,Kraus:1998hv,Cvetic:1999xp}  are explicit realizations of these phenomena. 
In particular, these are first black holes whose critical exponents were all computed, and this was the first time where dynamic critical phenomena have been ever discussed in the context of black hole physics.

In this talk, I will consider finite temperature criticality, not quantum criticality. Then, we will consider the finite temperature version of AdS/CFT. In this case, a finite temperature gauge theory is dual to a black hole in the AdS space. In the gravity side, a black hole appears since a black hole is a thermal system. Due to the Hawking radiation, a black hole has a notion of temperature. So, for our problem, above all things we need a black hole with a second-order phase transition. Is there any? 
\smallskip
\begin{itemize}

\item The simplest gravity system is pure gravity, and the solution is the Schwarzschild-AdS black hole.%
\footnote{I consider black holes with {\it planar horizon} only not black holes with compact horizon.}
 The gauge theory dual is the ${\cal N}=4$ super-Yang-Mills (SYM) theory. But the ${\cal N}=4$ SYM is conformal, so there is no phase transition. We need a more complicated system. 

\item The next simplest system is the Einstein-Maxwell system, and the solution is the Reissner-Nordstr\"{o}m-AdS black hole. But the system still has no phase transition. Adding a charge is not enough. Probably this is due to the no-hair theorem. According to the theorem, a black hole often has a few parameters, mass, angular momentum, and charge. Given these quantities, the black hole solution is unique. In order to have a phase transition, we need a multiple number of solutions. 

\item What we can do is to add a scalar. Usually, the existence of a scalar does not affect a black hole solution from the no-hair theorem. But the theorem is not entirely true for the AdS space. So, add a scalar to have different kinds of black holes. At least two interesting systems are known: they are R-charged black holes and holographic superconductors. And we study both.%
\footnote{See Refs.~\citen{Franco:2009if,Aprile:2009ai,Buchel:2009mf,Iqbal:2010eh,Herzog:2010vz} which study different systems.}

\end{itemize}
\smallskip
\noindent
Then, the AdS/CFT dictionary tells the mapping between gravity (``bulk") fields and field theory (``boundary") operators:


\begin{equation*}
\begin{array}{lcl}
\mbox{\it Bulk field} && \mbox{\it Boundary operator} \\
\mbox{graviton } h_{\mu\nu} & \leftrightarrow &  \mbox{energy-momentum tensor } T^{\mu\nu} \\
\mbox{Maxwell field } A_\mu & \leftrightarrow & \mbox{current } J^\mu \\ 
\mbox{scalar field } \Psi & \leftrightarrow & \mbox{scalar operator } \calO
\end{array}
\end{equation*}

\noindent
So, field theory ingredients contain $T^{\mu\nu}, J^\mu$, and $\calO$. There is such a one-to-one correspondence because bulk fields act as sources of boundary operators. 

We study two systems, holographic superconductors and R-charged black holes. You may ask why we should study both. This is because they belong to different universality classes. A holographic superconductor is dual to some kind of superconductor. The static universality class of holographic superconductors is just the conventional Ginzburg-Landau (GL) one. The $d=5$ R-charged black hole is dual to the ${\cal N}=4$ SYM at a finite chemical potential, and the static universality class is rather unconventional. They also belong to different dynamic universality classes. Holographic superconductors belong to model A, and R-charged black holes belong to model B in the classification of Hohenberg and Halperin \cite{hohenberg_halperin}.

\section{Static and dynamic critical phenomena} \label{sec:critical}

In a second-order phase transition, thermodynamic quantities diverge as powers of the correlation length, and the power is the critical exponent. Traditionally, there are 6 exponents.%
\footnote{
These exponents are defined as follows (for ferromagnets):
The specific heat: $C_H \propto |\eT|^{-\alpha}$, 
the spontaneous magnetization: $m \propto |\eT|^{\beta}~(T<T_c)$, 
the magnetic susceptibility: $\chi_T \propto |\eT|^{-\gamma}$,
the critical isotherm: $m \propto |h|^{1/\delta}  (T=T_c)$,
the correlation function $(T \neq T_c)$: $G(r) \propto e^{-r/\xi}$, 
the correlation function $(T=T_c)$: $G(r) \propto r^{-d_s+2-\eta}$, and 
the correlation length: $\xi  \propto |\eT|^{-\nu}$.
Here, $\eT:= (T-T_c)/T_c$, and $d_s$ denotes the number of {\it spatial} dimensions.
\label{fnote:exponents_def}
} 
As an example, consider the standard GL theory:
\be
f = c(\del_i m)^2 +  \frac{1}{2} a m^2 + \frac{1}{4} b m^4 + \cdots -m h~.
\label{eq:GL_ferro}
\ee 
The GL theory has the familiar Higgs-like potential, and the mass term is proportional to temperature: $a = a_0 (T-T_c) + \cdots (a_0>0)$. So, at high temperatures, the potential is minimum at the origin $m=0$, no magnetization. But at low temperatures, the system has nonzero magnetization $m \neq 0$. From this free energy, the GL theory has the following exponents:
\be
(\alpha, \beta, \gamma, \delta, \nu, \eta) = \left(0, \frac{1}{2}, 1, 3,  \frac{1}{2}, 0\right)~.
\label{eq:GL_exponents}
\ee
For example, the spatially homogeneous solution of \eq{GL_ferro} is given by $\del_m f |_{h=0} = am+bm^3 = 0$, so $m \propto \sqrt{-a/b} \propto \sqrt{T_c-T}$. 
This determines $\beta=1/2$. We will see that some black holes have the same values as the GL theory whereas some black holes do not have the values.

These exponents are not all independent, and they have to satisfy {\it static scaling relations}:
\be
\alpha + 2\beta+\gamma =2~, \quad
\gamma = \beta(\delta-1)~, \quad
\gamma = \nu (2-\eta)~, \quad
2-\alpha = \nu d_s~.
\label{eq:scaling}
\ee
The AdS implementations should satisfy these relations (except the last one, the hyperscaling relation, which often fails.)

In the dynamic case, the relaxation time $\tau$ also diverges as $\tau \propto \xi^z$, which is known as the critical slowing down. The exponent $z$ is the {\it dynamic critical exponent}. The details of the dynamic exponent depend on {\it dynamic universality classes}. The dynamic universality class depends on additional properties of the system which do not affect the static universality class. In particular, conservation laws play an important role to determine dynamic universality class. A conservation law forces the relaxation to proceed more slowly. As a consequence, even if two systems belong to the same static universality class, they may not belong to the same dynamic universality class. 

Typical macroscopic variables are conserved charges and order parameters. So, I am talking of the relaxation time of these variables. The dynamic universality classes were classified by Hohenberg and Halperin \cite{hohenberg_halperin}: they are known as {model A, B, C, H, F, G, and J. Since a conservation law plays an important role, the dynamic universality classes are classified whether the order parameter is conserved or not, and whether there are the other conserved parameters. Our primary focus is model A and B. For these models, only the order parameter matters. So, we will consider the relaxation of the order parameter. 

Let us consider a simple example to understand the critical slowing down. Suppose an order parameter $m$ satisfies the following equation of motion:
\be
\frac{dm (t)}{dt} = -\Gamma \del_m f 
\label{eq:dynamic_example_1}~.
\ee
Equation~(\ref{eq:dynamic_example_1}) tells that the time-dependence of the order parameter is controlled by the return force, which is given by the derivative of the free energy $f$. This is essentially the same as the damped harmonic oscillator. So, imagine a particle rolls the potential. Using the free energy for the GL theory (\ref{eq:GL_ferro}), the equation of motion becomes $dm/dt = - a \Gamma m$. 

The equation of motion is easily solved, and the solution is the exponential decay $m \propto e^{-t/\tau}$ with the relaxation time $\tau^{-1} = a \Gamma$. But $a$  vanishes at the critical point, so $\tau$ diverges at the critical point. Namely, the relaxation time diverges since the potential becomes flat at the critical point. Parametrize the divergence by the exponent $z$ using the correlation length: $\tau \propto \xi^z$. For the GL theory, $\xi \propto |T-T_c|^{-1/2}$, so $\tau^{-1} \propto \xi^{-2}$, namely $z=2$.
It is often convenient to use momentum space. In momentum space, the equation of motion takes the form ($\omega$: frequency)
\be
\omega = -i a \Gamma~.
\label{eq:slowing_down}
\ee 

We will consider only model A and B below. 
For model A, the order parameter is not conserved. In this case, the situation is the same as the above example, and $z=2$.

For model B, the order parameter is conserved. A conserved charge satisfies the diffusion equation $\del_t m - D\del_i^2m=0$, where $D$ is the diffusion constant. In momentum space, one has the dispersion relation ($q$: wave number)
\be
\omega = -iDq^2~.
\label{eq:diffusion}
\ee
Comparing \eq{slowing_down} with \eq{diffusion}, one gets $\Gamma \propto q^2$, and $D \propto a$.
But $a=0$ at the critical point, so $D=0$ at the critical point. We will use this fact later. Write the relaxation time in terms of the correlation length, and write the $q$-dependence as $(\xi q)$: 
$\tau^{-1} \sim a q^2 \sim \xi^{-2}q^2 \sim (\xi q)^2/\xi^4$. 
In this sense, $z=4$. 

More generally, the dynamic exponent is given by $z=2-\eta$ (model A) or $z=4-\eta$ (model B) using the static exponent $\eta$ (anomalous exponent). These are known as {\it dynamic scaling relations}. For the GL theory, $\eta=0$ so that dynamic scaling relations reduce to the above values. 
Part of our job is to find AdS counterparts of model A and B.

\section{Critical phenomena in the AdS/CFT duality} 

\subsection{Holographic superconductors}

First, let us discuss holographic superconductors. 
A holographic superconductor is a solution of gravity coupled with $U(1)$ gauge field and a complex scalar $\Psi$:
\be
S = \int d^{d+1}x \sqrt{-g} \left[ R - 2\Lambda - \frac{1}{4} F_{ab}^2  
- |(\nabla_a - iq A_a)\Psi|^2 -V(|\Psi|) \right]~.
\label{eq:H^3_action}
\ee 
There are two branches for the solution. At high temperatures, the solution is the standard charged black hole with no scalar. But this is not true at low temperatures. The solution becomes unstable and is replaced by a charged black hole with a scalar field. So, this scalar field $\Psi$ corresponds to the order parameter $\calO$ for the phase transition. However, analytic solutions are not known. So, people usually use numerical computations or use some approximations to solve the theory. 

The claim is that this corresponds to some kind of superconductor. In fact, the charge conductivity diverges for the black hole, and there is some sign of energy gap.  
This model corresponds to an extreme type II superconductor. For type II, magnetic field can penetrate the superconductor in the form of vortex. The vortex solutions have been constructed in this model \cite{Albash:2009ix,Albash:2009iq,Montull:2009fe,Maeda:2009vf}.%
\footnote{For the magnetic response, see also Refs.~\citen{Hartnoll:2008kx,Maeda:2008ir,Maeda:2010br}.}


For holographic superconductors, the analytic solution is unknown at low temperatures, but the analytic solution is known at high temperatures. It is just the standard Reissner-Nordstr\"{o}m black hole. So, we approach from high temperature. This gives a further simplification. In this regime, the scalar $\Psi$ decouples from the rest, gravity and electromagnetic perturbations. So, it is enough to solve the $\Psi$ equation. 
Then, we analyze the resulting equation of motion both analytically and numerically.

The system has the following static critical exponents \cite{Maeda:2009wv}%
\footnote{Some exponents have been computed previously \cite{Hartnoll:2008vx,Herzog:2008he,Maeda:2008ir,Horowitz:2008bn}.}:
\be
(\alpha, \beta, \gamma, \delta, \nu, \eta) = \left(0, \frac{1}{2}, 1, 3,  \frac{1}{2}, 0\right)~.
\ee
They are just the conventional values (\ref{eq:GL_exponents}). Holographic superconductors are first black holes whose critical exponents were all computed. 

For the dynamic case, the order parameter is the condensate $\calO$ dual to the scalar $\Psi$. It is not a conserved quantity, so the system must belong to model A. We indeed show that $z=2$. 

Our results are independent of dimensionality which is typical for mean-field results. It is natural to expect mean-field behaviors since one expects that fluctuations are suppressed at large-$N$ so that mean-field results become exact.

\subsection{R-charged black holes}

Let us discuss the next example, R-charged black holes. The $d=5$ R-charged black hole is dual to the ${\cal N}=4$ SYM at a finite chemical potential $\mu$. The action is again  gravity coupled with a gauge field and a scalar $H$:
\be
   S = \int d^{5}x \sqrt{-g}\, \left[~R - \frac{L^2}{8}\, H^{4/3}\, F_{ab}^2
  - \frac{1}{3}\, \frac{(\nabla_a H)^2}{H^2}
  + 2 V(H)~\right]~.
\label{eq:reduced-action-1charge}
\ee
In that sense, the action is similar to holographic superconductors, but there are some differences. 

The analytic solution is known for this system. So, one can compute thermodynamic quantities. You can show a second derivative of the free energy, the charge susceptibility $\chi:=\del\rho/\del\mu$ diverges at some point. 
On the other hand, first derivatives of the free energy such as the entropy density are regular there. So, this is a second-order transition.

The system has the following static exponents \cite{Cai:1998ji,Cvetic:1999rb,Maeda:2008hn,Buchel:2010gd}:
\be
(\alpha, \beta, \gamma, \delta, \nu, \eta) = \left(\frac{1}{2}, \frac{1}{2}, \frac{1}{2}, 2, \frac{1}{4}, 0\right)~.
\ee
From thermodynamic quantities, one can easily compute first 4 critical exponents. Two critical exponents are related to correlation functions, but correlation functions are much harder to compute \cite{Buchel:2010gd}. Anyway, the exponents take unconventional values compared with the GL theory (\ref{eq:GL_exponents}).%
\footnote{Even though they have unconventional values, they satisfy scaling relations (\ref{eq:scaling}) except the hyperscaling relation. Also, one should regard them as unconventional but mean-field results since we are taking the large-$N$ limit.} 
So, this system is not described by the conventional GL Hamiltonian. In other words, AdS/CFT implies {\it the existence of a new fixed point which has been never observed in a standard field theory.}

Let us consider the dynamic case. Since the charge susceptibility diverges, the black hole has a singular behavior in the charge density. So, the charge density $\rho$ is the order parameter. Since $\rho$ is a conserved quantity, the system must be model B. From our previous discussion, one expects the diffusion constant $D$ vanishes. The diffusion constant can be derived from the current-current correlator by Kubo formula, and there is a standard AdS/CFT technique to compute it. I cannot discuss the procedure in detail here, but the boundary current is dual to the bulk Maxwell field. So, essentially we solve the bulk Maxwell field. One can show that $D=0$ at the critical point \cite{Maeda:2008hn} and $z=4$ \cite{Buchel:2010gd} as in model B. 

To summarize, some black holes do obey the theory of critical phenomena. I would like to stress that it is a priori not obvious. We are talking of the critical phenomena of black holes, not a usual condensed-matter system. In the presence of gravity, the usual notion of statistical mechanics does not hold in general. For example, there is no stable thermal equilibrium. The AdS/CFT duality claims that AdS black holes are equivalent to usual statistical systems such as gauge theories. So, in this case we expect that black holes behave as in the theory of critical phenomena, and we indeed saw this. 

We considered two examples. For holographic superconductors, the static universality class is the conventional GL theory one, and the dynamic universality class is model A. For R-charged black holes, the static universality class is unconventional, and the dynamic universality class is model B. 

\section*{Acknowledgements}
I would like to thank Kengo Maeda and Takashi Okamura for collaboration and useful discussions. 
I would like to thank the Yukawa Institute for Theoretical Physics at Kyoto University and the organizers of NFQCD2010 for their hospitality and for a stimulating environment. This work was supported in part by the Grant-in-Aid for Scientific Research (20540285) from the Ministry of Education, Culture, Sports, Science and Technology, Japan.

%

\end{document}